\documentstyle[aps,multicol,eqsecnum,epsf]{revtex}
\begin{document}
\draft
\title{Static Friction between Elastic Solids due to Random Asperities}
\author{J. B. Sokoloff}
\address{Physics Department and Center for Interdisciplinary Research on
Complex Systems, Northeastern University, Boston, MA 02115.}
\date{\today }
\maketitle

\begin{abstract}

Several workers have established that the Larkin domains for two three 
dimensional nonmetallic elastic solids in contact
with each other at a disordered interface are
enormously large, implying  
that there should be negligible static friction per unit area in the 
macroscopic solid limit.
 The present work argues that the
fluctuations in the heights of the random asperities at the interface 
that occur in the Greenwood-Williamson model can account
for static friction. 
\end{abstract}

\begin{multicols}{2}
\narrowtext

It is well known that one 
must apply a minimum force (known as static friction) in order to get two 
solids, which are in contact, to slide relative to each other. 
Several workers have provided 
evidence, however, that there might be no static friction for 
nonmetallic crystalline surfaces, 
incommensurate with each other. Aubry showed this for the weak potential 
limit of one dimensional Frenkel-Kontorova model[1], and recently 
He, et. al.[2] and Muser and Robbins[3] have shown for weakly interacting 
two dimensional 
incommensurate interfaces that the force of static friction per unit area 
falls to zero as 
$A^{-1/2}$ in the thermodynamic limit, where A is the interface area. 
Even solids made 
of identical materials are incommensurate if their crystalline axes are 
rotated with respect to each other. 
Disorder, however, can pin contacting solids, just as it pins sliding 
charge density waves[4,5] 
and vortices in a superconductor[6]. 
Recently, it has been 
shown that Larkin domains (i.e., domains over which the solids are able to distort to accommodate 
the disorder at the interface) 
for contacting three dimensional elastic solids 
are enormously large compared to typical solid sizes[7-9], 
implying that the force of static friction per unit area due to interface 
disorder should also fall off as $A^{-1/2}$ in the thermodynamic limit. 
In Refs. 2 and 3 it was proposed that the presence of 
a submonolayer film of mobile molecules at the interface is a requirement 
for the occurrence 
of static friction between incommensurate surfaces, which appears to 
imply that 
clean interfaces are frictionless.
In the present work, it is argued that the Greenwood-Williamson 
model (GW)[10,11] predicts the existence of sparsely spaced higher 
than average asperities at the interface which 
interact more strongly with the second surface than with each other. 
Since these are in the strong pinning limit, they will exhibit static friction. 

Here, scaling methods, like those used by Fisher 
for charge density waves (CDW)[12], are used to study static friction for 
non-smooth interfaces. 
This can be accomplished by minimizing the potential energy of the solid in 
contact with a rigid disordered substrate at z=0 with 
respect to the size of a Larkin domain[4], which is expected to give 
qualitatively correct results for the problem of two elastic disordered 
solids in contact. Given that the energy density 
of the elastic solid is given approximately by
$$(1/2)E'|\nabla {\bf u}|^2+V({\bf r})\delta(z), \eqno (1)$$
where $E'$ is an effective Young's modulus and 
$V({\bf r})$ is the potential per unit area of the disordered substrate and 
${\bf u}({\bf r})$ is the local displacement of the solid, the 
energy of a single Larkin domain is given by
$$E=(1/2)L'L^2 E'\alpha [|\nabla_t' {\bf u'}|^2/L^2+$$
$$|\partial {\bf u'}/\partial z'|^2/L'^2]-V_0 a L, \eqno (2)$$
where a is a local length scale (e.g., a lattice constant), L is the width 
and L' is the height of the domain, $|\nabla_t' {\bf u'}|^2=
|\partial {\bf u'}/\partial x'|^2+|\partial {\bf u'}/\partial y'|^2$, 
where we assume that the local displacement {\bf u} varies on length scales 
L and L' in the x and y and 
the z directions, respectively. That is, we assume that ${\bf u}(x,y,z)$ 
has the form {\bf u'}(x',y',z'), where the function u' varies by an amount 
of the order of local length scales (e.g., a lattice constant) when x', y' and 
z', defined by (x',y',z')=(x/L,y/L,z/L'), each vary by an amount of order unity.
Here, $V_0$ is a typical value of the potential per unit area. When 
Eq. (2) is minimized 
with respect to L' one finds that $L'\approx L$ and the energy per unit 
volume within a Larkin domain at the interface is given by
$$E/L^3\approx [(1/2)E' |\nabla' {\bf u'}|^2-V_0 a]/L^2, \eqno (3)$$
(where we use the average value of $|\nabla' {\bf u'}|^2$ here) 
whose absolute minimum occurs for infinite L (more correctly L comparable to 
the interface length) for $E'|\nabla' {\bf u'}|^2>V_0 a$, implying that 
the static friction per 
atom decreases as the reciprocal of the square root of the surface area. 
These scaling arguments apply equally well to 
disorder on the sub-asperity level due to atomic level defects and to 
the multi-micron length scale level due to asperities. 
When applied at the sub-asperity level, 
they show that the contact force between asperities should be proportional 
to the square root of area 
of contact. On the multi-asperity length scale, 
the atoms are replaced by asperities and the defects by contacting asperities.

If the sliding solid has dimensions normal to the interface much 
smaller than those along it, L' can only be as large as the
 thickness. 
Minimization of Eq. (3) with respect to L with L' fixed at the thickness 
shows that the 
Larkin length is comparable to the film thickness. 
Thus, the interface consists of many Larkin domains. Since the 
pinning force 
scales with the number of Larkin domains at the 
interface, the force of static friction per unit area 
is non-zero value, in the large solid limit.  

Fisher[12] has shown that above the critical dimensions of 4, 
charge density 
waves are not pinned for typical impurity strengths, but fluctuations 
in the impurity concentration and strength lead to pinning. 
(The critical dimension for two solids in contact is 3, as seen above.)   
Consider the effect of fluctuations in the defect concentration 
for thick solids, by dividing the solid into boxes of length L 
and examining the percentage of blocks at the interface of 
sufficiently large defect concentration to be in the 
"strong pinning" regime, where the substrate force 
on each block 
dominates over the inter-block elastic forces.  
Consider the parameter $h\approx V_0/\alpha a^2$, where 
$\alpha$ is the interatomic force constant, a is a lattice 
spacing and $V_0$ is the strength of the potential due to a defect acting 
on an atom. Let $n_c=c'L^2$ be the number of defects within a particular 
block and $c'$ (where $c'>c$, where c is the average defect concentration 
for the interface), the defect concentration strong enough 
for the block to be considered a strong block.  
Then the ratio of the interaction of a typical strong 
block with the substrate to $\alpha a^2$ is $h(c'L^2)^{1/2}$. The interface 
area surrounding each strong block is the total interface area A divided by 
the number of strong blocks at the interface, $PA/L^2$, 
where P is the probability of a particular block being a strong one. Then 
$L^2/P$ is the interface area surrounding each block and the 
typical length L' over which the elastic interaction between two strong blocks 
acts is its square root, $L'=L/P^{1/2}$. 
Then the total elastic energy associated with each strong block is 
the product of the volume per strong block=$(L')^3$ times the elastic energy 
density, which is proportional to $|\nabla u|^2$ [which scales as $(L')^{-2}$]
 or $L'$. The criterion for a block to be a strong one is 
$h(c'L^2)^{1/2}>>L'$, or $h>>(c' P)^{-1/2}$. Since $c' P<1$, this violates 
our previous assumption that $h<<1$, implying that such 
fluctuations cannot result in strong pinning. There are also fluctuations 
in the locations of the points of contact within the defect potential wells 
within each Larkin domain, it too does not lead to static friction[13].

Defect strength fluctuations, however, can lead to static friction, as 
we shall see.  On the multi-asperity scale, the 
surface asperity height distribution 
is likely to produce such fluctuations for sufficiently large surfaces even 
those that are quite smooth. This is the situation for the GW
model[10,11,13,14], in which there are elastic spherical asperities 
on a surface with an exponential or Gaussian height distribution 
in contact with a rigid substrate. The GW model is generally accepted to 
be a correct way to account for Anonton's laws in most cases[11], 
especially for relatively light loads.
Volmer and Nattermann's discussion of static 
friction[14] is not qualitatively different from that of Ref. 10. 
In the GW model, the 
total contact area is of the order of
$$A_c=2\pi\sigma bN\int_h^{\infty} ds \phi (s) (s-h), \eqno (4)$$
where $\phi (s)$ is the distribution of asperity heights z, where $s=z/\sigma$,
  where $\sigma$ is a length
 scale associated with the height distribution, and h is the 
the ratio of the distance of the lower part of the bulk part of the 
sliding solid, from the surface in which it is in 
contact, to $\sigma$, b is the radius of curvature of an asperity and N 
is the number of asperities 
above a certain size, independent of whether they are in contact[10,11,13,14]. 
The interaction of a single asperity with the substrate 
is equal to the product of the contact area and a shear strength for the 
interface. Actually, to be consistent 
with our scaling arguments and Refs. 2 and 3 we should assume that the 
friction force on a single asperity is proportional to the square root of 
the asperity contact area, but when this was done, the results were not 
changed qualitatively.

The energy of the interface consists of two parts. One part is the single 
asperity energy, 
which consists of the interaction energy of an asperity 
with the substrate plus the elastic energy cost necessary for 
each asperity to seek its minimum energy, neglecting its elastic interaction 
with other asperities, which is independent of the asperity density. 
The second part is the elastic interaction between 
asperities within the same solid, which depends on the asperity density.
In order to determine these energies, 
let us model the interaction of the $\ell^{th}$ asperity with the substrate 
by a 
spherically symmetric harmonic potentials of force constant $\alpha_{\ell}$. 
Assume that 
in the absence of distortion of the solid, the $\ell^{th}$ asperity  
lies a distance  
${\bf \Delta}_{\ell}$ from the center of its potential 
well. Let ${\bf u}_{\ell}$ be the displacement of the $\ell^{th}$ 
asperity from its initial position.  
We use the usual elastic Green's function tensor of the medium at a distance 
r from the point at which a force is applied at the interface, 
but for simplicity, we approximate it by 
the simplified form $G(r)=(E'r)^{-1}$, where E' is 
Young's modulus[15]. Then the 
equilibrium conditions on the u's are 
$${\bf u}_{\ell}=(E'a)^{-1}\alpha_{\ell} ({\bf \Delta}_{\ell}-{\bf u}_{\ell})+
\sum_j (E'R_{\ell,j})^{-1}
\alpha_j ({\bf \Delta}_j-{\bf u}_j), 
\eqno (5)$$
where a is a parameter of the order of the size of the asperity. 
To lowest order in the interasperity interaction, approximate the solution
for ${\bf u}_{\ell}$ is
$${\bf u}_{\ell}={\bf u}_{\ell}^{0}+[1+(E'a)^{-1}\alpha_{\ell}]^{-1}
\sum_j (E'R_{\ell,j})^{-1}
\alpha_j ({\bf \Delta}_j-{\bf u}^0_j), 
\eqno (6)$$
where ${\bf u}_{\ell}^0$ is the zeroth order approximation (i.e., the solution 
to Eq. (5) neglecting the second term on the right hand side of the equation). 
Since the contacting asperities are randomly distributed over the interface, 
we can estimate the second term (i.e., the summation over j) on the right hand 
side of Eq. (6) by its root mean square (RMS) average which is estimated by 
integrating the square of the summand over the position of the $j^{th}$ 
asperity, which is in contact with the substrate, over its position and 
multiplying by the density of asperities in contact with the substrate 
$\rho$. Since the angular integrals only give a factor of order unity, 
we need only evaluate the integral over the magnitude of $R_{\ell,j}$, giving 
an RMS value of the sum over $R^{-1}$ of order $[\rho ln(L/a)]^{1/2}$ 
where here L is 
the length of the interface and a is the asperity size. For $L\approx 1cm$ 
and $a\approx 10^{-6}cm$, $[ln(L/a)]^{1/2}$ 
is of order unity. The energy of the system can 
be written as 
$$(1/2)\sum_j \alpha_j |{\bf \Delta}_j-u_j|^2+$$
$$(1/2)E'\sum_j\int d^3 r [|\nabla {\bf G}({\bf r})|
(\alpha_j |{\bf\Delta}_j-{\bf u}_j)|]^2 \eqno (7)$$
It follows from Eqs. (5-7) that the two lowest order nonvanishing 
terms in an 
expansion of the energy of the system in powers of $\rho^{1/2}$ are the 
zeroth and first order ones. 
Since the 
shearing of the junction at the area of contact of two asperities involves 
the motion of two atomic planes realtive to each other, the distance over 
which the contact potential varies must be of the order of atomic distances. 
Then, if we denote the width of the asperity contact potential well by b, 
of the order of atomic spacings, we must choose a typical value 
for $\alpha$ such that $\alpha b$ is of the order of the shear rupture 
strength of the asperity contact junction. Thus, $\alpha>>E'a$. 
Zeroth order in the asperity density in Eq. (7) is of the order of 
$\alpha\Delta^2$, where $\alpha$ is a typical value of $\alpha_j$, 
and $\Delta$ is a typical value of $\Delta_j$. 
The term linear in $\rho^{1/2}$ is easily shown to be of the order of 
$E'a^2\Delta^2 \rho^{1/2}$ to zeroth order in $E'a/\alpha$. 
Since it depends on $\rho$ it 
represents an interaction energy between the asperities.  
Then, the mean inter-asperity interaction is proportional to  
the square root of the number of contacting asperities per unit surface 
area, given by
$$\rho (h)=(N/A)\int_h^{\infty} ds \phi (s). \eqno (8),$$
where A is the total surface area and N is the total number of asperities 
whether in contact with the substrate or not. The integral in 
Eq. (4) divided by the integral in Eq. (8), which is proportional to the 
contact area per asperity and the square root of the integral in 
Eq. (8) are plotted as a function of the load, which is given in this 
model by
$$F_L=(4/3)E'N(b/2)^{1/2}\sigma^{3/2}\int_h^{\infty} ds \phi (s) (s-h)^{3/2}, 
\eqno (9)$$
in Fig. 1. 

\begin{figure}
\centerline{
\vbox{ \hbox{\epsfxsize=4.5cm \epsfbox{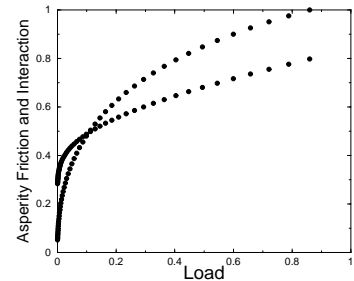}}
       \vspace*{1.0cm}
        }
}
\caption{The curve which is lower at the right is a plot of the integral in 
Eq. (4) divided 
by the integral in Eq. (8) and the curve which is higher on 
the right is a plot of the square root of the integral in Eq. (8) versus 
the integral in Eq. (9). 
All quantities are dimensionless.}
\label{Fig1}
\end{figure}
 \noindent A Gaussian distribution is assumed here for $\phi (s)$ (i.e., 
$\phi (s)=(2\pi)^{-1/2}e^{-s^2/2}$).
Since the square root of Eq. (8) drops to zero 
in the limit of vanishing load, whereas Eq. (4) divided by Eq. (8) 
approaches a nonzero 
value, this implies that the interface will approach the strong pinning 
regime (i.e., the regime in which the asperity-substrate interaction 
dominates over the inter-asperity interaction) in the limit of vanishing load. 

Let us now give sample numerical values for 
some of the quantities which occur in the 
application of the GW model to this problem. For example, 
for a typical plastic, Young's modulus is $4.0\times 10^{9}N/m^2$ and the 
shear rupture 
strength is $3.5\times 10^{7}N/m^2$[17]. Following Ref. 11, we choose $\sigma=
2.4\times 10^{-4}mm$ and $b=6.6\times 10^{-2} mm$, and assume that there 
is a density of $4.0\times 10^3$ asperities/$mm^2$. 
Then by performing the integrals 
in Eqs. (4), (8) and (9), we find that for $F_L/A=3.98\times 10^{-4} N/mm^2$, 
where A is the apparent area of the interface, the total contact area divided 
by A is $3.03\times 10^{-5}$, and the contact area per asperity from 
the ratio of Eqs. (4) and (8) is $2.44\times 10^{-5} mm^2$. 
Also, $\rho (h)^{1/2}$, which is equal to 
the square root of Eq. (8) is $1.11 mm^{-1}$. The mean interasperity 
interaction 
force is approximately equal to the derivative of the first order term in 
$\rho^{1/2}$ given above Eq. (8) with respect to $\Delta$ 
or $E'a^2\rho (h)^{1/2}\Delta$, where a is taken as the square 
root of the mean contact area per asperity divided by $\pi$. 
The mean strength of the force acting on an asperity, 
due to the solid with which it is in contact, will be estimated by the product 
of its contact area and the shear rupture strength $E_r$. 
Then, the condition for the latter 
quantity to dominate over the asperity-asperity interaction,  
$E_r\pi a^2>E' 4\pi a^2\rho^{1/2}\Delta$ or 
$E_r/E'>4\rho (h)^{1/2}\Delta$, is easily satisfied by the above 
calculated quantities since the right hand side is $4\times 10^{-7}$ and 
the left hand side is $8.75\times 10^{-3}.$ 

Although for higher loads the system appears to move towards the 
"weak pinning" limit, 
the latter conclusion is most likely incorrect because it does not take 
into account the fact that the distribution of asperity heights contains 
asperities which are much higher than average. These asperities will be 
compressed much more than a typical asperity, making the friction force on 
them considerably larger than average. Since the probability of such 
asperities occurring is relatively small, however, they will be typically 
far apart, putting them in the strong pinning limit. For example, the 
probability of the ratio of an asperity height to $\sigma$ 
being greater than a value $h_L$ is
$$P(h_L)=\int_{h+h_L}^{\infty} ds \phi (s), \eqno (10)$$
whose mean height and hence contact area is proportional to 
$$P(h_L)^{-1}\int_{h+h_L}^{\infty} ds \phi (s) (s-h). \eqno (11)$$
These two quantities are plotted in Fig. 2.
It is seen that even for $h_L$ only equal to 1/2 (corresponding 
to an asperity height 
equal to $(1/2)\sigma$), Eq. (11) 
remains larger than the square root of Eq. (10). 

Although it has been argued here that the GW model 
predicts the occurrence of a sufficiently dilute concentration of 
asperities with stronger than average forces acting on them due to the second 
solid to consider the asperities to be essentially uncorrelated, this still 
does not necessarily guarantee that there will be static 
friction, since it has been argued that even for uncorrelated asperities static 
friction will only occur if the asperities exhibit multistability[9,16]. The 
condition for multistability to occur at an interface[9], namely that the 
force constant due to the asperity contact potential be larger than 
that due to the elasticity of the asperity ($\approx E'a$), however, will be 
satisfied, as discussed earlier.

\begin{figure}
\centerline{
\vbox{ \hbox{\epsfxsize=4.5cm \epsfbox{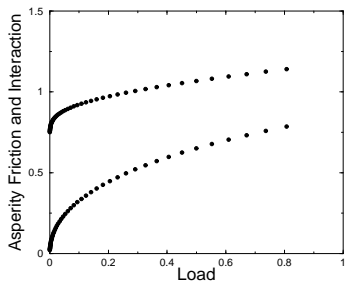}}
       \vspace*{1.0cm}
        }
}
\caption{Eq. (11) (the higher curve) and the square root of Eq. (10) 
(the lower curve) are 
plotted versus the load Eq. (9) 
divided by $(4/3)E(b/2)^{1/2}\sigma^{3/2}$. All quantities are dimensionless.}
\label{Fig2}
\end{figure}

In conclusion, although at first sight it appeared that 
arguments based on Larkin domains indicate that the disorder at an interface 
between two nonmetallic elastic solids in contact would not result in static 
friction, when one takes into account the fluctuations in asperity height 
that occur in models such as the GW model, there will always exist 
asperities with greater than average height to put them in the 
"strong pinning regime," resulting in static friction. For light loads,   
even typical height asperities (as opposed to height fluctuations) can 
easily be in the "strong pinning limit."

\acknowledgments

I wish to thank the Department of Energy (Grant DE-FG02-96ER45585).

\end{multicols}{2


\begin{references}
\bibitem {} J. B. Sokoloff, Phys. Rev. B23, 1992 (1981). One way of 
understanding this is that if there is static friction, the sliding velocity 
of the solid will only be nonzero if a force above the force of static 
friction is applied to the body. Alternatively, we can view this as implying 
that the force of friction approaches a nonzero value as the center of mass 
velocity approaches zero. The arguments in this reference and Ref. 9 tell us 
that there must be multistability for this to occur.
\end{references}
\end{document}